\documentclass[a4paper]{article}
\usepackage{interspeech2010,amssymb,amsmath,epsfig}
\usepackage{amsmath}
\usepackage{amssymb,amsfonts,textcomp}
\usepackage{url}
\usepackage{array}
\ninept
\def\reg{{\rm\ooalign{\hfil
     \raise.07ex\hbox{\scriptsize R}\hfil\crcr\mathhexbox20D}}}

\title{ Should Corpora be Big, Rich, or Dense?}

\newcommand\T{\rule{0pt}{2.6ex}}

\makeatletter
\def\name#1{\gdef\@name{#1\\}}
\makeatother
\name{{\em Greg P. Kochanski$^1$, Chilin Shih$^2$, Ryan Shosted$^2$}}
\address{$^1$The University of Oxford, UK\\
  $^2$The University of Illinois, Urbana-Champaign\\
{\small \tt greg.kochanski@phon.ox.ac.uk, cls@illinois.edu, rshosted@illinois.edu}}

\begin{document}
\maketitle
{\tiny The reference for this paper is:
Greg P. Kochanski, Chilin Shih, Ryan Shosted, ``Should Corpora be Big, Rich, or Dense?'',
presented at \emph{New Tools and Methods for Very-Large-Scale Phonetics Research},
University of Pennsylvania, January 28-31, 2011.
\url{http://www.ling.upenn.edu/phonetics/workshop/index.html}}

To see a world in a grain of sand \newline
And a heaven in a wild flower,\newline
Hold infinity in the palm of your hand\newline
And eternity in an hour.

\ \ \ \ \ \ \ \ \ \ \ \ \ \ \ \ \ \ \ William Blake - Auguries of
Innocence

\begin{abstract}
In this paper, we ask what properties makes a large corpus more or less
useful. We suggest that size, by itself, should not be the ultimate
goal of building a corpus. Large{}-scale corpora are considered
desirable because they offer statistical stability and rich variation.
But this rich variation means more factors to control and evaluate,
which can limit the advantages of size. We discuss the use of
multi{}-channel data to complement large{}-scale speech corpora. Even
though multi{}-channel data may limit the scale of a corpus (due to the
complex and labor{}-intensive nature of data collection) they can offer
information that allows us to tease apart various factors related to
speech production.
\end{abstract}
\noindent{\bf Index Terms}: corpora, experimental, linguistics, speech, articulation, large

\section{Why use a Large Corpus?}

Not too long ago, the concept of a large linguistic corpus
didn{\textquotesingle}t exist; neither did the infrastructure necessary
to build and maintain such a corpus. Recently, speech technology has
opened up the possibility of conducting large experiments. Consider an
enthusiastic human communicator who makes 200 hours of phone calls per
month. Digitized at 16 bits, 16 kHz over a 90-year lifetime, this
amounts to just 25 Terabytes, for a lifetime storage cost of
$\approx${}US\$10,000.\footnote{Assuming November 2010 storage costs, no future
price reductions, and a disk lifetime of 10 years.} Given a suitable
speech recognition system, that lifetime of data could be transcribed
if a comparable amount of money were spent on computers and
electricity. We are approaching the point where we can now
investigate an entire language, rather than a small sample. Since the
costs of such enormous corpora are suddenly within the realm of
possibility, we ask how they should be designed.

In the past, speech corpora have been small; increasing the size was
intended to increase statistical power. If one is counting linguistic
items\footnote{I.e.\ the frequency with which an word (or other
linguistic items) occurs in a text. Or, more generally, the frequency
of a particular word (phone, phrase, accent, {\dots}) combination in a
particular context.}, 1000 examples ($N=1000$) are much more informative
than one, because they allow you to estimate the frequency of the word
precisely\footnote{The confidence intervals and statistical
significance of frequency measurements can be modeled by Poisson
statistics, where the fractional accuracy of a frequency measurement is
$N^{-1/2}$, where $N$ is the number of occurrences of the items. So,
$N=1000$ occurrences allows you to measure an item's
frequency within 3\%.}, whereas a single example gives only the crudest
possible idea of how common the word is. Similarly, a single
measurement of an acoustic property means little, because from it we
learn nothing about variability. Ten samples allow us to measure
variability in one dimension; one hundred or a thousand samples allow
us to come up with multidimensional correlations.

In principle, more repetitions of a word will allow for a more precise
measurement of the average properties of a sound, but the benefits of
repetition taper off beyond $N=1000$.  Currently, we
don{\textquotesingle}t know of two theories of speech variation that
can be differentiated by measurements at this level of precision. It is
possible that theories of speech variation will never be this precise
because language is not part of the Newtonian
``clockwork universe'', and some of
the observed variation may be inherent to a stochastic communication
system.

In natural speech (or near approaches to it), the frequency distribution
often follows Zipf's law \cite{Zipf1935, Zipf1949}: There are a
few items in a corpus with very high frequency, more items with lower
frequencies, but most items have a very low frequency. One example is
the distribution of words: 5\% of an English text corpus is
``the'', but most words are more like ``haggard'', with frequencies
near 0.0001\%. \ Any particular word like ``haggard'' may not even appear in a
corpus of less than a million words, even though such words (as a
group) form much of the corpus. For applications that need a good
representation of infrequent events, such as an automatic speech
recognition system, it is crucial to train the systems using a very
large corpus. This ensures correct recognition of infrequent words or
unusual combinations of sounds in a variety of dialects.

We can define a boundary between
``small'' and ``large'' corpora by asking whether
the most common items occur often enough ($N > 1000$) to
allow for good measurements. In a small corpus, examples of all items
are scarce; in a large corpus at least the most common items are
sufficiently represented. The next natural step is a huge corpus,
where most items have $N > 1000$. Large corpora are
appearing; huge corpora (except for phones) are still rare (Table~1).
However, even the biggest current audio corpora, like the BNC \cite{Coleman:this}
are just entering the ``large'' category if one wishes to study how one word affects another.

\textbf{Table 1:} Large and Huge phonetic corpora.\\
\begin{tabular}{p{2.5cm}|p{2cm}|p{2cm}}
\hline
Research on: & How big is a ``large'' corpus? & \dots a ``huge'' corpus?\\\hline
Individual phones \T & $> 10^3$ words &
$> 10^5$ words\\\hline
Triphones \T &
$> 10^5$ words &
$> 2\cdot 10^6$ words\\\hline
Triphones with prosody\footnote{
Assuming several prosodic factors, such as stress, focus, distance from speaker to listener, and noise level.
} &
$> 10^6$ words \T &
$> 4\cdot 10^9$ words\\\hline
Individual words \T &
$> 3\cdot 10^5$ words  &
$> 10^9$ words\\\hline
Word bi{}-grams \T &
$> 10^7$ words &
$> 10^{15}$ words\\\hline
\end{tabular}

If one starts with a minimally large corpus, because of
Zipf's law there will be only a few items whose
frequencies can be measured precisely. If we make the corpus bigger,
this charmed circle of items with $N > 1000$ will slowly
expand. \ So, very large corpora help studies of rare items -- and
recall that most linguistic items are rare. As can be seen in Table
1, one would need to expand the corpus by factors of hundreds,
thousands, or even millions to be able to study an entire language,
instead of studying merely its most frequent items.

\section{Natural Speech vs. Experiments}

The extreme amount of data needed for a huge corpus is a consequence of
the rarity of many linguistic items (ie. Zipf's law,
interpreted broadly). But this is not a logical necessity, merely a
description of the language that people produce in daily life.
Techniques like sociolinguistic interviews (cf.\ \cite{Labov2006}) and map
tasks (cf.\ \cite{McAllister1990}) are useful to boost the frequency of a
selected group of words while the speaker(s) still produce speech that
is reasonably natural.

These approaches are steps along a continuum towards a laboratory
experiment, where the speech is under the
experimenter's control, and normally rare words and
word combinations can be induced to occur as frequently as desired.
So, for some purposes, laboratory experiments are far more efficient
than a large corpus analysis. If a conclusion can be reached by
examining a small fraction of the items in the whole language, and if
these items can be easily induced, then an experiment may be appropriate.

But, experiments have difficulties over and above the the possibility of
phonetic differences between speech in a formal experiment and more
natural situations (cf.\ \cite{Kochanski2007}). An experiment
(and the associated analysis) is often set up to decide between two
possible hypotheses carefully chosen by the experimenter, based on the
results of previous studies. When the null hypothesis is rejected,
people may mistakenly assume the alternative is proven. This logic
follows Sherlock Holmes' famous dictum
``When you have eliminated the impossible, whatever
remains, however improbable, must be the truth.''
\cite{Doyle1927}. While misleading, the dictum is not exactly wrong in the
strict sense that the truth must be somewhere among whatever remains.
However, Doyle (or Holmes?) was wrong to suggest that this was a useful
way to solve difficult problems. It fails because when we apply it,
our notion of ``\dots whatever remains\dots '' is
limited by the human imagination, but the correct answer isn't.

The universe presents answers that people find hard to believe or
imagine, so it is hard to design an experiment that anticipates them.
In contrast, large speech corpora offer variations of language use and
speech production that may be unexpected and hard to imagine. With
large natural corpora, it is possible to break out of the limitations
of one's own imagination when one sees something
unexpected.

\section{Limits of Large Corpora}

In addition to their advantages, large corpora have disadvantages, too.
Expanding a corpus often introduces extra factors into a statistical
analysis. A small corpus might be very uniform: it might be acquired
in a short time, in a restricted location, with a carefully defined
dialect, in a uniform speaking style, under controlled recording
conditions. Large corpora often allow some of these factors to vary,
either for practical reasons, or intentionally, as a way to explore
their effect. And, with each new factor, one should allocate some of
the data towards understanding the effect of the factor.

An (extreme) example can illustrate this point. Imagine a small corpus
of English collected in Singapore, then double its size by adding
American English. Singapore English is heavily influenced by its
proximity to Chinese: it has different pronunciation, intonation,
rhythm (\cite{Low2000}, though see \cite{Loukina2009}) and word frequency. Any prosody
research using the expanded corpus would probably be best done by
partitioning the corpus into two halves, and analyzing each half
separately. As a result, the expanded corpus will provide no better
description of the prosody of Singapore English than the
original.\footnote{Of course, the hypothetical enlarged corpus will
allow dialect{}-to{}-dialect comparisons for whichever prosodic
properties can be measured on the original corpus. However, we would
only be able to measure and publish those comparisons if the corpus
reliably separates speakers of the two dialects. Many do not, and
fall back upon self{}-reporting and/or geographic information (e.g.\ the
British National Corpus).} This is an example where certain
questions remain unanswerable, no matter how many dialects one adds to
the corpus\footnote{Under some conditions, with a large and diverse
corpus, the research questions can be broadened from (e.g.)
``properties of a dialect'' to
``properties of the language'' when
more dialects are added. However, this should only be done in cases
where it is reasonably clear that these average properties are relevant
to real individuals who speak the language. For instance,
``small'' and ``wee'' are equivalent words in two
British dialects, and British English as a whole might use
``wee'' 0.1\% of the time (Google statistics for ``wee child'' vs.\ ``small child''),
but there may not be any actual individuals who use those two words interchangeably at
the population average rate.}.

Sometimes, if there are confounds amongst the extra factors, they do not
even yield interesting comparisons. For instance, one can imagine a
corpus intended to sample the speech that the average British person
would hear in the 1970s. It might be comprised of informal middle
class speech in the local dialect and formal, RP speech from the BBC.
Interpreting the difference between the two types of speech would be
hindered because one would not know whether to attribute a difference
to social class or to the formality of the presentation. Similar
confounds between factors are common in speech data: the word pairs in
a corpus are constrained by grammar, and the phone pairs in a word are
limited to those present in the lexicon.\footnote{For instance, in a
coarticulation experiment, one would like to be able to form all
combinations of sounds to see how each sound affects all others. But
most combinations are either unfamiliar to most speakers, or can only
be formed across word boundaries.}

So, though size may have benefits, extra, uncontrolled factors often
present in a larger corpus will erase some of the advantage: rich
variation of a corpus is not necessarily an advantage unless the goal is to study
variation. To an extent, one should think of a corpus in terms of the
density of data per factor: the ratio between the size of a corpus and
the number of combinations of relevant factors. If there is not
enough data to support each factor, it will be impossible to find the
best{}-fitting (possibly true), multi{}-factor explanation, no matter
the size of the corpus. In other words, the design of the corpus can
be more important than its size, especially as we move through the
range of large, into huge corpora.

\section{Multi{}-Channel Data}

Multi{}-channel data allow us to increase the data density of a corpus;
such data can be used to complement controlled experiments and large,
speech{}-only databases. Of course, having multiple data channels is
nothing new to speech scientists, because any speech signal can be
interpreted as a group of related signals, e.g. the power in
various frequency bands may each be interpreted as separate
signals.\footnote{As in a MFCC front end for a speech recognition
system.}

By ``multi{}-channel corpora'' we
mean corpora where the acoustics of speech are recorded along with
other related signals. Data that can be recorded alongside speech
acoustics include articulatory movement (Electromagnetic Articulography, ultrasound, fiberoscopy),
linguopalatal contact (EPG), airflow and pressure, muscle activity
(EMG), as well as facial and hand gestures.\footnote{
Part-of-speech annotation and other annotation might also count for something here, though such
annotation carries relatively little information.}
In contrast to the
large{}-scale speech corpus which are ``horizontally
rich''
we view
multi{}-channel data as
``vertically'' rich\footnote{
Horizontally = large in terms of time;
Vertically = large in terms of the number of measurements per time point.
Data data are typically plotted on the y{}-axis against time.}.

Acoustic signals we record tell us something about the state of the oral
articulators, but it is well{}-known that they render incomplete
information. For instance, multiple articulatory configurations can
generate virtually the same acoustic signal \cite{Schroeder1967,Mrayati}.
This means, for instance, that one cannot deduce the state of
the articulators from 100 milliseconds of a speech
signal.\footnote{Note that with longer speech signals, it is
sometimes possible to use the idea that the motions of the articulators
must be smooth and continuous to remove some ambiguities. See \cite{Schroeter1994}.
}

The ambiguity can become harder to resolve when one tries to deduce
features of the language that are deeper than articulatory positions.
For example, when an English speaker emphasizes a word, they may use
a longer duration. But long durations are also associated with final
lengthening and focus. So (absent other information), the case of a
long syllable is ambiguous. Likewise, loudness can be associated with
focus, emphasis, or low vowels, so observation of loudness alone cannot
tell you the prosodic function. Fant put it neatly:
``The translation from speech wave back to
articulation is to some extent restricted by the existence of
compensatory forms of articulation\dots A deeper insight into the
potentialities of this aspect of the physiological interpretation of
spectrograms must rely on extensive correlative
work'' \cite[p.\ 209]{Fant1960}.

In some cases, the function of a gesture can be deduced by comparing
several aspects of an acoustic signal. But humans experience richer
communication in person than over the telephone, so there is good
reason to believe that face, hand, and arm gestures are an important
part of our communication. They may carry information of their own in
addition to disambiguating the acoustics. To pick a trivial example,
one cannot easily convey a shrug over the telephone. That information
is either lost to the listener, or the speaker adapts to the
communication channel and packages the information in some other form. 

Multi{}-channel data can be especially important when there are
trade{}-off relationship between different factors. For example,
while duration, loudness, and $f_0$ are recognized (across language) as
important acoustic correlates of stress or emphasis, a speaker
doesn't need to use all factors at the same time to
convey linguistic meaning. This might be implemented as a trade{}-off
relationship where if a speaker lengthens the duration for emphasis,
changes in loudness or $f_0$ would be unnecessary. Given such a
trade{}-off, any one measurement (e.g.\ duration) would show large
amounts of variation across emphasized syllables, but the correct
combination of multiple properties would add up to some gestalt of
emphasis with much less variability\footnote{Strong trade{}-off
relationships (to the extent that they exist) are important because
they indicate that variability in certain combinations of acoustic
parameters is linguistically unimportant. Absent knowledge of the
trade{}-off, this variability would likely be interpreted as a
difference in meaning or function.}. 

Also, the articulatory{}-acoustic mapping is nonlinear (cf.\ \cite{Schroeder1967,Stevens2000})
This means that (for instance) a 1mm closing gesture can be easily
perceived in the confines of a narrow airway, but may be acoustically
undetectable in an open airway. However, if one has
formant information along with articulatory information, the formant
information can provide a detailed view of the articulation near
closure, and the articulatory measurements will constrain hypotheses
about what may be going on when the airway is open.\footnote{One
might reasonably ask ``why do the articulatory details
matter when the airway is nearly open if it has no acoustic
consequences?'' First, your conversation partner
may be watching you, so jaw opening may count as a facial gesture.
Second, even for telephone speech, the width of opening is related to
the velocity of the following closure, which may have audible
consequences.}

Overall, adding data beyond audio measurements into a corpus can add
substantial information that is not otherwise available. From the
perspective of data density, this data brings along a minimum of extra
factors because it is a simultaneous view of the exact same instance of
a word. Contrast this with a horizontal expansion of a corpus: you
can easily bring in new instances of the same word, but the new
instances come without any reason to believe that they are equivalent
to the instances you already have.\footnote{I ndeed, if there is a
relevant trade{}-off relationship that involves non{}-acoustic data,
then one might well falsely conclude that two instances did not have
equivalent meanings or functions.} They are uttered in new conditions
(typically we must introduce new factors to describe these conditions\footnote{
	Having metadata about the utterances will clearly help, but it should
	be noted that metadata derived from the audio is not strictly new,
	independent information.
	}),
or simply uttered differently because of unexplained
instance{}-to{}-instance variation. When you add a second instance of
a word to a corpus, you cannot determine whether it is identical to the
first word without spending some of the data's
explanatory power. In effect, one must introduce new factors that
describe the differences between pairs of potentially identical words
and new questions to answer\footnote{Every pair of words comes with the
implicit question ``Are these words
linguistically/functionally/phonologically equivalent or
not?''}. On the other hand, if you add multiple,
simultaneous measures of a related signal, all for the same word, each
measure corresponds to the exact same word you started out with. There
is no question regarding the identity of the word, it is merely being
viewed from a different angle.\footnote{There will, typically, be some
data spent to determine the relationship between acoustic and
articulatory measurements. However, that is often more like an
initial calibration, and one does not have a new increment of
uncertainty with each new instance.}

\section{Conclusion}

It is generally agreed that multiple recordings of a given item will
allow us to better understand variation, i.e. by revealing tendencies
in the data from which we can make statistical inference. It follows
that we should collect large numbers of items in order to make better
predictions that generalize to the population. Corpus linguistics, as
traditionally conceived, suggests that more observations of a
phenomenon enable us to better understand the phenonmenon. While size
generally helps, it is not always the case, and the design details can
be very important.  In some cases, a larger corpus raises more
questions, and the increase in questions can cancel out the increase in size.
Especially in cases where trade{}-offs are
important or interpretation is ambiguous, multi{}-channel corpora with
a relatively small number of items may have a comparable value to much
larger acoustic{}-only corpora.

\section{Acknowledgments}

We thank John Coleman for comments.
Greg Kochanski appreciates support from the UK ESRC via RES-062-23-2566, RES-062-23-1172, and RES-062-23-1323.

\eightpt
\bibliographystyle{IEEEtran}

\begin{thebibliography}{10}

\bibitem[3]{Coleman:this}
Coleman, J., Liberman, M., Kochanksi, G., Burnard, L., and Yuan. J.
`Mining a Year of Speech'', also submitted to this conference.


\bibitem[7]{Doyle1927}
Doyle, Sir Arthur. (1927). The Adventure of the Blanched Soldier, in
The Casebook of Sherlock Holmes.


\bibitem[10]{Schroeder1967}
Schroeder, M. R. (1967). ``Determination of the vocal tract shape from measured formant frequencies'',
J.\ Acoustic.\ Soc.\ Am.\ 41, pp. 1283--1294.

\bibitem[13]{Fant1960}
Fant, Gunar. (1960). Acoustic Theory of Speech Production. Mouton \& Co,
The Hague, Netherlands.

\bibitem[6]{Kochanski2007}
Kochanski, G., and Orphanidou, C. (2007) ``Testing the
Ecological Validity of Repetitive Speech''
Proceedings of the International Congress of Phonetic Sciences (ICPhS
XVI), Saarbrücken, Germany.
\url{http://www.icphs2007.de/conference/Papers/1632/1632.pdf}

\bibitem[9]{Loukina2009}
Loukina, A., Kochanski, G., Shih, C., Keane, E., and Watson, I. (2009)
``Rhythm measures with language-independent segmentation'',
	        \emph{Proceedings of the 10th Annual Conference of the International
		        Speech Communication Association} (Interspeech 2009). ISSN 1990-9772
			        Brighton, UK, 7--10 September 2009, pp 1531--1534.


\bibitem[8]{Low2000}
Ling, Low Ee, Grabe, Esther, and Nolan, Francis. (2000).
``Quantitative characterizations of speech rhythm:
Syllable{}-timing in Singapore English''. Language
and Speech, 43 (4), pp.\ 377-401.

\bibitem[4]{Labov2006}
Labov, William. (2006). The Social Stratification of English in New York
City. Cambridge University Press.

\bibitem[5]{McAllister1990}
McAllister, J., Sotillo, C., Bard E.G., and Anderson, A.H. (1990).
``Using the map task to investigate variability in
speech,'' Occasional paper, Department of Linguistics,
University of Edinburgh.

\bibitem[11]{Mrayati}
Mrayati, M., Carre, R., and Guerin, B. (1988) ``Distinctive regions and modes: a new theory of speech production'',
\emph{Speech Communication} 7(3), p. 257--286.

\bibitem[12]{Schroeter1994}
Schroeter, Juergen and Sondhi, Mohan. (1994). Techniques for estimating
vocal{}-track shapes from the speech signal. IEEE Transactions on
Speech and Audio Processing, Vol 2, No 1, pp.\ 133-150.

\bibitem[14]{Stevens2000}
Stevens, Kenneth. (2000). Acoustic Phonetics. The MIT Press. 

\bibitem[1]{Zipf1935}
Zipf, George K. (1935). The Psychobiology of Language.  Houghton{}-Mifflin. 

\bibitem[2]{Zipf1949}
Zipf, George K. (1949). Human Behavior and the Principle of Least
Effort. Cambridge, MA: Addison{}-Wesley.
\end{thebibliography}

\end{document}